\documentclass[journal,10pt,twocolumn,oneside, peereview]{IEEEtran}
\usepackage{enumerate}
\usepackage{float}
\usepackage{amssymb,amsmath,amsxtra,amsfonts,amsthm,mathtools}
\usepackage{graphicx}
\usepackage{bigints}
\usepackage{fancyhdr}
\usepackage{float}
\usepackage{times}
\usepackage{wrapfig}
\usepackage{xcolor,colortbl}
\usepackage{soul}
\usepackage{array,multirow,makecell}
\usepackage{subfig}
\usepackage{nopageno}
\usepackage{tikz}
\usepackage{glossaries}
\usepackage{hyperref}
\usepackage[switch,columnwise]{lineno} 
\usepackage{mathrsfs}
\usepackage{lineno}
\usepackage[
backend=biber,
style=numeric,
sorting=ynt
]{biblatex}

\usepackage{amsthm}

\begin{document}
\title{\textbf{Reconfigurable FPGA-Based Solvers For Sparse Satellite Control}}

\author{
  Anis Hamadouche,\,\, 
  Yun Wu,\,\,
  Mathini Sellathurai,\,\,
  Andrew M.\ Wallace,\,
  and\,
  Jo\~ao F.\ C.\ Mota%
  \IEEEcompsocitemizethanks{
    \IEEEcompsocthanksitem
    \IEEEcompsocthanksitem
    Anis Hamadouche, Mathini Sellathurai, Yun Wu, Andrew M.\ Wallace, and Jo\~ao F.\ C.\ Mota
    are with the School of Engineering \& Physical Sciences, Heriot-Watt University, Edinburgh EH14 4AS,
    UK. (e-mail: \{ah225,y.wu,a.m.wallace,j.mota\}@hw.ac.uk).
  }
}

\maketitle
\thispagestyle{plain}
\pagestyle{plain}

\begin{abstract}
\textbf{This paper introduces a novel reconfigurable and power-efficient FPGA (Field-Programmable Gate Array) implementation of an operator splitting algorithm for Non-Terrestial Network's (NTN) relay satellites model predictive orientation control (MPC). Our approach ensures system stability and introduces an innovative reconfigurable bit-width FPGA-based optimization solver. To demonstrate its efficacy, we employ a real FPGA-In-the-Loop hardware setup to control simulated satellite dynamics. Furthermore, we conduct an in-depth comparative analysis, examining various fixed-point configurations to evaluate the combined system's closed-loop performance and power efficiency, providing a holistic understanding of the proposed implementation's advantages.}
\end{abstract}

\IEEEpeerreviewmaketitle
\newcommand\scalemath[2]{\scalebox{#1}{\mbox{\ensuremath{\displaystyle #2}}}}
\section{INTRODUCTION}

In Non-Terrestial Networks (NTN), relay satellites can be used to transmit signals over horizons, avoiding interference or blockages that might occur in densely populated or urban areas. Proper orientation ensures that the satellite's antennas or communication modules face the intended direction, minimizing the chances of interference from unintended sources. In NTN, redundancy and reliability can be achieved by having multiple relay satellites in place that can provide backup options in case one satellite fails or is temporarily out of position. Although this redundancy enhances the overall reliability of the network, timely and proper satellite orientation becomes even more challenging.

Precise relay satellites orientation is a cornerstone for an effective NTN and space optical communication. It ensures that communication links are established accurately, efficiently, and reliably, meeting the demands of modern space communication needs. For instance, an efficient and a reliable NTN optical wireless communication (OWC) requires very precise alignment because laser beams have a much narrower beamwidth compared to RF signals and even a small misalignment can lead to communication breakdowns. Therefore, consistent and accurate orientation mechanisms ensure that satellites can quickly re-establish lost links or switch between communication partners, maintaining network uptime.

FPGAs, or Field-Programmable Gate Arrays, offer a unique advantage in digital system design due to their reconfigurability. This allows for custom hardware implementations tailored to specific tasks, leading to optimized performance and reduced power consumption. In the context of satellite control and communication, precision is paramount. Even minor inaccuracies in calculations can lead to significant deviations in satellite orientation, potentially causing mission failures or at the very least, increased fuel/power consumption for course corrections. When combined with sparse and/or maximum hands-off control strategies, it can offer unparalleled advantages, especially in systems requiring communication efficiency, minimal intervention, minimal energy, and adaptation~\cite{nagahara2016discrete, ikeda2021maximum}. Sparse control minimizes signaling overhead which can lead to efficient satellite communication where bandwidth is premium~\cite{barforooshan2019sparse}. When deployed on FPGA, these controls enable systems to operate efficiently, autonomously, and adaptively - attributes that are indispensable for battery-powered satellites with communication systems grappling with resource constraints and ever-changing environments.

In this work, we implement a first-order operator splitting algorithm on an FPGA and we leverage the flexibility of arbitrary bit-width assignment to control the power consumption of the FPGA fabric while controlling a simulated satellite dynamics in real-time. We then perform a comparative study between different bit-width and fraction-width selections in light of Model Predictive Control (MPC) performance and power efficiency. 

In Section~\ref{section:problem}, we delve into the mathematical intricacies underlying MPC. Moving forward, Section~\ref{section:mainresults} introduces the advanced generalised WLM-ADMM algorithm. Our practical application is demonstrated in Section~\ref{section:experiment}, where we implement an approximate proximal-gradient algorithm—a specific instance of the WLM-ADMM algorithm—on a ZCU106 FPGA evaluation board. This section further showcases a FPGA-In-the-Loop simulation of satellite dynamics, executed with varying precision levels. Finally, Section~\ref{section:conclusion} concludes the paper, offering valuable insights and shedding light on potential avenues for future research.

\section{PROBLEM FORMULATION}
\label{section:problem}
A linear dynamical system can be described by a linear differential equation as follows
\begin{align}
    \dot{x}(t) &= Ax(t) + Bu(t),\quad t\geq 0, \quad x(0) \in \mathbb{R}^n \label{StateSpaceContinuous}
\end{align}
where $x(t) \in \mathbb{R}^n$ is the state and the control vector is given by $u(t) \in \mathbb{R}^p$. The matrices $A, B$ have dimensions $n \times n$, $n \times p$, respectively. The closed-form solution of the differential equation~\eqref{StateSpaceContinuous} can be obtained analytically and it is given by
\begin{equation}
    x(t) = e^{A(t-t_0)} x(t_0) + \int_{t_0}^t e^{A(t-\tau)}B u(\tau) d\tau,\quad t \geq 0. \label{CCSol}
\end{equation}
The control problem is to design a control signal $u(t)$, for $t \geq 0$, that forces the system~\eqref{StateSpaceContinuous} to manifest some desired dynamics, or equivalently, to force the trajectory $x(t)$~\eqref{CCSol} to follow some specified trajectory.
Let $u_{x[k]} = [u_{x[k]}(0), u_{x[k]}(h),\dots,u_{x[k]}((N-1)h)]^\top$ be the solution of the following optimal control problem
\begin{align}
\label{DOPTl0}
    &\underset{u_{x[k]} \in \mathbb{R}^{p\times (N_c-1)}}{\min} \sum_{i=0}^{N_c-1} \mathcal{L}_u(u_{x[k]}[i]),\notag\\
    \text{s.t.}&\quad [A_d^{N-1}B_d\quad A_d^{N-2}B_d \dots B_d] u_{x[k]} + A_d^N x(0) = 0,\\
    &\quad            \|u_{x[k]}\|_{\infty}\leq u_{\max},\notag
\end{align}
where
\begin{equation}
    (A_d, B_d,C,D) = \left(e^{Ah}, \int_0^h e^{At}B dt,C,D\right).
\end{equation}

and $x(T) = x[N] = 0$. 

With initial condition $x[k]$ instead of $x(0)$, MPC is obtained by feeding the computed control vector at instant $k$ to the system under control and measure the updated state. The latter is used as the new initial condition of the optimal control algorithm as follows
\begin{equation}
\begin{split}
    x[k+1] = A_d x[k]+B_d [1\quad 0 \dots 0] u_{x[k]}\\
    y[k] = C x[k]+D [1\quad 0 \dots 0] u_{x[k]}\\
\end{split}
\end{equation}
Based on the the state space model, future states are predicted using future (planned) control sequence as follows
\begin{align}
    x[k+1|k] &= A_d x[k]+B_d u_{x[k]}[0]\\
    x[k+2|k] &= A_d^2 x[k]+A_dB_d u_{x[k]}[0]+B_d u_{x[k]}[1]\\
    x[k+3|k] &= A_d^3 x[k]+A_d^2B_d u_{x[k]}[0]+A_dB_d u_{x[k]}[1]\notag\\&+B_d u_{x[k]}[2]\\
    \vdots\\
    x[k+N_p|k] &= A_d^{N_p} x[k]+A_d^{N_p-1}B_d u_{x[k]}[0]+\dots\notag\\&+A_d^{N_p-N_c}B_d u_{x[k]}[N_c-1].
\end{align}
Likewise, the predicted output trajectory is given by
\begin{align}
    y[k+1|k] &= CA_d x[k]+CB_d u_{x[k]}[0]\\
    y[k+2|k] &= CA_d^2 x[k]+CA_dB_d u_{x[k]}[0]+CB_d u_{x[k]}[1]\\
    y[k+3|k] &= CA_d^3 x[k]+CA_d^2B_d u_{x[k]}[0]\notag\\&+CA_dB_d u_{x[k]}[1]+CB_d u_{x[k]}[2]\\
    \vdots\notag\\
    y[k+N_p|k] &= CA_d^{N_p} x[k]+CA_d^{N_p-1}B_d u_{x[k]}[0]+\dots\notag\\&+CA_d^{N_p-N_c}B_d u_{x[k]}[N_c-1].
\end{align}
and $u_{x[k]}$ is the solution of the following MPC problem
\begin{align}
\label{MPC}
    \underset{u_{x[k]} \in \mathbb{R}^{p\times (N_c-1)}}{\min}& \sigma \sum_{i=0}^{N_c-1} \mathcal{L}_u(u_{x[k]}[i])+\sum_{i=0}^{N_p-1} \mathcal{L}_y(y[k+i|k],r[k+i]),\notag\\
    \text{s.t.}&\quad \Phi u_{x[k]} = Y-Fx[k],\\
    &\quad  \begin{bmatrix}-I\\I\\-\Phi\\\Phi\end{bmatrix}u_{x[k]} \leq \begin{bmatrix}-u_{\min}\\u_{\max}\\-y_{\min}+Fx[k]\\y_{\max}-Fx[k]\end{bmatrix}.\notag
\end{align}
where 
\begin{align}
    Y &= \begin{bmatrix}y[k+1|k]\\\vdots\\y[k+N_p|k]\end{bmatrix},\label{Eq:Y}\\
    \Phi &= \begin{bmatrix}
    CB& \dots& 0\\
    CAB& CB& \dots& 0\\
    CA^2B& CAB& \dots& 0\\
    \vdots& \ddots& \ddots& \vdots\\
    CA^{N_p-1}B& CA^{N_p-2}B& \dots& CA^{N_p-N_c}B\end{bmatrix},\label{Eq:Phi}\\
    F &= \begin{bmatrix}CA\\\vdots\\CA^{N_p}\end{bmatrix}\label{Eq:F}.
\end{align}
\section{MAIN RESULTS}
\label{section:mainresults}
In this section we present tho major theoretical contributions. A general framework for constrained MPC with sparsity-promoting cost functions as well as a general class of operator splitting algorithms to solve it.  

Sparse MPC is obtained when the following stage costs are used 
\begin{align}
&\mathcal{L}_u(u_{x[k]}) = |u_{x[k]}|^p,\quad p \in \{0,1\}\\
&\mathcal{L}_y(y[k+i|k],r[k+i]) = \notag\\& \frac{1}{2}(y[k+i|k]-r[k+i])^\top Q (y[k+i|k]-r[k+i]),
\end{align}
where $Q$ is a positive definite matrix. Substituting the stage costs, the objective function of problem~\eqref{MPC} becomes
\begin{align}
    &\underset{u_{x[k]} \in \mathbb{R}^{p\times (N_c-1)}}{\min} \sigma \sum_{i=0}^{N_c-1} |u_{x[k]}[i]|^p\notag\\&+\frac{1}{2}\sum_{i=0}^{N_p-1} (y[k+i|k]-r[k+i])^\top Q (y[k+i|k]-r[k+i]),
\end{align}
which can be written matrix form
\begin{align}
    \underset{u_{x[k]} \in \mathbb{R}^{p\times (N_c-1)}}{\min}& \sigma \|u_{x[k]}\|_p+\frac{1}{2}(Y-R_s)^\top Q (Y-R_s),
\end{align}
where $R_s$ is an $m\times N_p$ matrix resulting from $N_p$ stacking of the target vector $r[k]$, i.e.,  $R_s:=[r[k]\quad r[k]\quad \dots\quad r[k]]^\top$ assuming $r[k+i] = r[k]$ for $0\leq i< N_p$. Finally, substituting $Y = \Phi u_{x[k]}+Fx[k]$ we obtain
\begin{align}
    &\underset{u_{x[k]} \in \mathbb{R}^{p\times (N_c-1)}}{\min} \sigma \|u_{x[k]}\|_p\notag\\&+\frac{1}{2}(\Phi u_{x[k]}+Fx[k]-R_s)^\top Q (\Phi u_{x[k]}+Fx[k]-R_s)\notag\\&+\mathcal{I}_{\mathcal{C}}(M u_{x[k]}),
    \label{MPCQl1}
\end{align}
where
\begin{align}    
      \mathcal{I}_{\mathcal{C}}(u) = \begin{cases} 0, & x\leq \mathcal{N}\\ \infty, & x > \mathcal{N}\end{cases}
\end{align}
with inequalities are applied elementwise, and matrices $\mathcal{M}$, $\mathcal{N}$ are given by 
\begin{align}
\mathcal{M} = \begin{bmatrix}-I\\I\\-\Phi\\\Phi\end{bmatrix},\quad
\mathcal{N} = \begin{bmatrix}-u_{\min}\\u_{\max}\\-y_{\min}+Fx[k]\\y_{\max}-Fx[k]\end{bmatrix}.
\end{align}
Defining $z = [z_0\quad z_1]^\top$ and $\mathcal{A} = [I \quad M^\top]^\top$, we can write \eqref{MPCQl1} as a decoupled composite optimization problem 
\begin{align}
    \underset{u_{x[k]} \in \mathbb{R}^{p\times (N_c-1)}}{\min}& \frac{1}{2}(\Phi u_{x[k]}+Fx[k]-R_s)^\top Q (\Phi u_{x[k]}+Fx[k]\notag\\&-R_s)+\sigma \|z_0\|_p+\sigma \mathcal{I}_{\mathcal{C}}(z_1),\\
    \text{s.t.}&\quad \mathcal{A} u_{x[k]} = z\notag
    \label{ComMPCQl1}
\end{align}
Choosing $g(u_{x[k]}) = \frac{1}{2}(\Phi u_{x[k]}+Fx[k]-R_s)^\top Q (\Phi u_{x[k]}+Fx[k]-R_s)$ and $h(z) = \sigma\|z_0\|_p+\mathcal{I}_{\mathcal{C}}(z_1)$, the WLM-ADMM iterates are given by
\begin{subequations}
\label{WLMADMMComMPCQl1}
    \begin{alignat}{3}
    &u^{k+1} = \underset{u}{\arg\min}\quad \frac{1}{2}(\Phi u+Fx[k]-R_s)^\top Q (\Phi u+Fx[k]-R_s)\notag\\&+\frac{1}{2}\|u -\Lambda_{1_k}^{-1}\gamma_{1_k}\|_{\Lambda_{1_k}}^2&\\
    &z^{k+1} = \underset{z}{\arg\min}\quad \sigma \|z\|_p+ \mathcal{I}_{\mathcal{C}}(z_1)+\frac{1}{2}\|z -\Lambda_{2_k}^{-1}\gamma_{2_k}\|_{\Lambda_{2_k}}^2&\\
    &v^{k+1} = v^k+(\mathcal{A}u^{k+1}-z^{k+1}).&
    \end{alignat}
\end{subequations}
where 
\begin{subequations}
\label{W$^2$LM-ADMM1.2Vars}
    \begin{alignat}{3}
    & \Lambda_{1_k} = \frac{1}{\lambda_u} \mathcal{A}^\top L  \mathcal{A} + M_u&\\
    & \gamma_{1_k}(u^k,z^k,v^k) = M_u u^k+\frac{1}{\lambda_u}\mathcal{A}^\top L  (z^k-v^k)&\\
    & \Lambda_{2_k} = \frac{1}{\lambda_z} L + M_z&\\
    & \gamma_{2_k}(u^{k+1},z^k,v^k) = M_z z^k+\frac{1}{\lambda_z}L  (\mathcal{A}u^{k+1}+v^k).&   
    \end{alignat}
\end{subequations}
For diagonal matrix $\Lambda_{2_k} = \operatorname{diag}\{\alpha_1,\dots,\alpha_{n},1,\dots,1\}$, and when $\|\cdot\|_p = \|\cdot\|_1$ we obtain
\begin{subequations}
    \begin{alignat}{3}
    &z^{k+1} = \begin{bmatrix} \mathcal{S}_{1/\alpha_i}\left([\gamma_{2_k}]_i\right) \\ \Pi_{\mathcal{C}}\left(\gamma_{2_k}\right)\end{bmatrix} = \begin{bmatrix} \mathcal{S}_{\sigma /\alpha_i}\left([\gamma_{2_k}]_i\right) \\ \operatorname{sat}_{\mathcal{N}}\left(\gamma_{2_k}\right)\end{bmatrix}&
    \end{alignat}
\end{subequations}
where $[\cdot]_i$ stands for the $i$-th vector component and $\mathcal{S}_{1/\alpha_i}$ is the (elementwise) \textit{soft thresholding} operation with $\sigma /\alpha_i$ threshold
\begin{align}
\mathcal{S}_{\sigma /\alpha_i}\left([\gamma_{2_k}]_i\right) = \begin{cases}
[\gamma_{2_k}]_i-\frac{\sigma }{\alpha_i}, & [\gamma_{2_k}]_i\geq\frac{\sigma }{\alpha_i}, \\
0, & |[\gamma_{2_k}]_i|<\frac{\sigma }{\alpha_i} \\
[\gamma_{2_k}]_i+\frac{\sigma }{\alpha_i}, & [\gamma_{2_k}]_i\leq-\frac{\sigma }{\alpha_i},
\end{cases}, \quad i = 1, \dots, n
\end{align}
When $\|\cdot\|_p = \|\cdot\|_0$ we obtain
\begin{subequations}
    \begin{alignat}{3}
    &z^{k+1} = \begin{bmatrix} \mathcal{H}_{\sqrt{2\sigma/\alpha_i}}\left([\gamma_{2_k}]_i\right) \\ \Pi_{\mathcal{C}}\left(\gamma_{2_k}\right)\end{bmatrix} = \begin{bmatrix} \mathcal{H}_{\sqrt{2\sigma/\alpha_i}}\left([\gamma_{2_k}]_i\right) \\ \operatorname{sat}_{\mathcal{N}}\left(\gamma_{2_k}\right)\end{bmatrix}&&
    \end{alignat}
\end{subequations}
where $\mathcal{H}_{\sqrt{2\sigma/\alpha_i}}$ is the (elementwise) \textit{hard thresholding} operation with $\sqrt{2\sigma/\alpha_i}$ threshold
\begin{align}
\mathcal{H}_{\sqrt{2\sigma/\alpha_i}}[\gamma_{2_k}]_i = \begin{cases}
[\gamma_{2_k}]_i, & |[\gamma_{2_k}]_i| >\sqrt{\frac{2\sigma}{\alpha_i}}, \\
0, & |[\gamma_{2_k}]_i|<\sqrt{\frac{2\sigma}{\alpha_i}} \\
\{0,[\gamma_{2_k}]_i\}, & |[\gamma_{2_k}]_i|=\sqrt{\frac{2\sigma}{\alpha_i}},
\end{cases}, \quad i = 1, \dots, n
\end{align}

From~\cite{hamadouche2023improved}, \eqref{WLMADMMComMPCQl1} can be written as a \textit{Moreau envelope} gradient descent:

\begin{subequations}
\label{Alg:Moreau-GD}
    \begin{alignat}{3}
    &u^{k+1} =  \gamma_{1_k} - \frac{1}{\lambda_u} \nabla \mathcal{M}_{\frac{1}{\lambda_u} g}(\gamma_{1_k}),&\\
    &z^{k+1} = \gamma_{2_k} - \frac{1}{\lambda_z} \nabla \mathcal{M}_{\frac{1}{\lambda_z} h}(\gamma_{2_k}),&\\
    &v^{k+1} = v^k+\mathcal{A}\gamma_{1_k}-\gamma_{2_k}- \frac{1}{\lambda_u}\mathcal{A}\nabla \mathcal{M}_{\frac{1}{\lambda_u} g}(\gamma_{1_k})\notag\\&+\frac{1}{\lambda_z}\nabla \mathcal{M}_{\frac{1}{\lambda_z} h}(\gamma_{2_k})&
    \end{alignat}
\end{subequations}

where 

\begin{subequations}
\begin{alignat}{3}
&\mathcal{M}_{\frac{1}{\lambda_u} g}(\gamma_{1_k}) = \underset{u}{\inf}\quad \frac{1}{\lambda_u} g(u)+\frac{1}{2}\|u -\Lambda_{1_k}^{-1}\gamma_{1_k}\|_{\Lambda_{1_k}}^2,&\\
&\mathcal{M}_{\frac{1}{\lambda_z} h}(\gamma_{2_k}) = \underset{z}{\inf}\quad \frac{1}{\lambda_z} h(z)+\frac{1}{2}\|z -\Lambda_{2_k}^{-1}\gamma_{2_k}\|_{\Lambda_{2_k}}^2&,
\end{alignat}
\end{subequations}

with $g(u) = \frac{1}{2}(\Phi u_{x[k]}+Fx[k]-R_s)^\top Q (\Phi u_{x[k]}+Fx[k]-R_s)$ and $h(z) = \sigma\|z_0\|_p+\mathcal{I}_{\mathcal{C}}(z_1)$.
\section{EXPERIMENTAL RESULTS}
\label{section:experiment}

In the context of our study, it is crucial to investigate both the power consumption and the resulting system performance under various configurations on the FPGA. For the sake of simplicity, we consider the unconstrained MPC problem and we choose $M_u = I$, $L = I$ and $M_z = 0$ in~\eqref{W$^2$LM-ADMM1.2Vars} to obtain the following approximate proximal-gradient descent algorithm (AxPGD) (according to~\eqref{Alg:Moreau-GD})
\begin{subequations}
\label{AxPGD}
    \begin{align}
    u^{k+0.5} &=  u^k - \frac{1}{\lambda_u} \nabla \mathcal{L}_y^{\epsilon_{\mathcal{L}_y}^k}(u^k),\label{AxPGD-x-update}\\
    u^{k+1}&=\mathcal{S}_{\sigma s}     \Big[u^{k} - s\big(\nabla \mathcal{L}_y(u^{k})+\epsilon_{\mathcal{L}_y}^{k}\big)\Big]\,.
    \end{align}
\end{subequations}
where $s = 1/\alpha_i = \frac{1}{\lambda_u} = 0.0002$ and the inexact $\epsilon_{\mathcal{L}_y}^k$-gradient $\nabla \mathcal{L}_y^{\epsilon_{\mathcal{L}_y}^k}(u^k)$ is defined as
\begin{equation}
    \nabla \mathcal{L}_y^{\epsilon_{\mathcal{L}_y}^k}(u^k) := \nabla \mathcal{L}_y (u^k)+ \epsilon_{\mathcal{L}_y}^k.
\end{equation}
with $\epsilon_{\mathcal{L}_y}^k$ being the approximation error that is associated with FPGA arbitrary precision. We choose $\sigma = 1.5$ for this experiment.

We used HLS design flow to generate the IP core of~\ref{AxPGD}. In order to validate the latter, we use the ZCU106 evaluation board to perform FPGA-In-the-Loop simulation after programming the PL fabric with the generated bitstream.

The figures below show both the orientation of an NTN's relay satellite, the power of the programmable logic (PL) and the control signal in a real-time FPGA-In-the-Loop (ZCU106) simulator. 

      \begin{figure}[!htb]
        \includegraphics[width=\columnwidth]{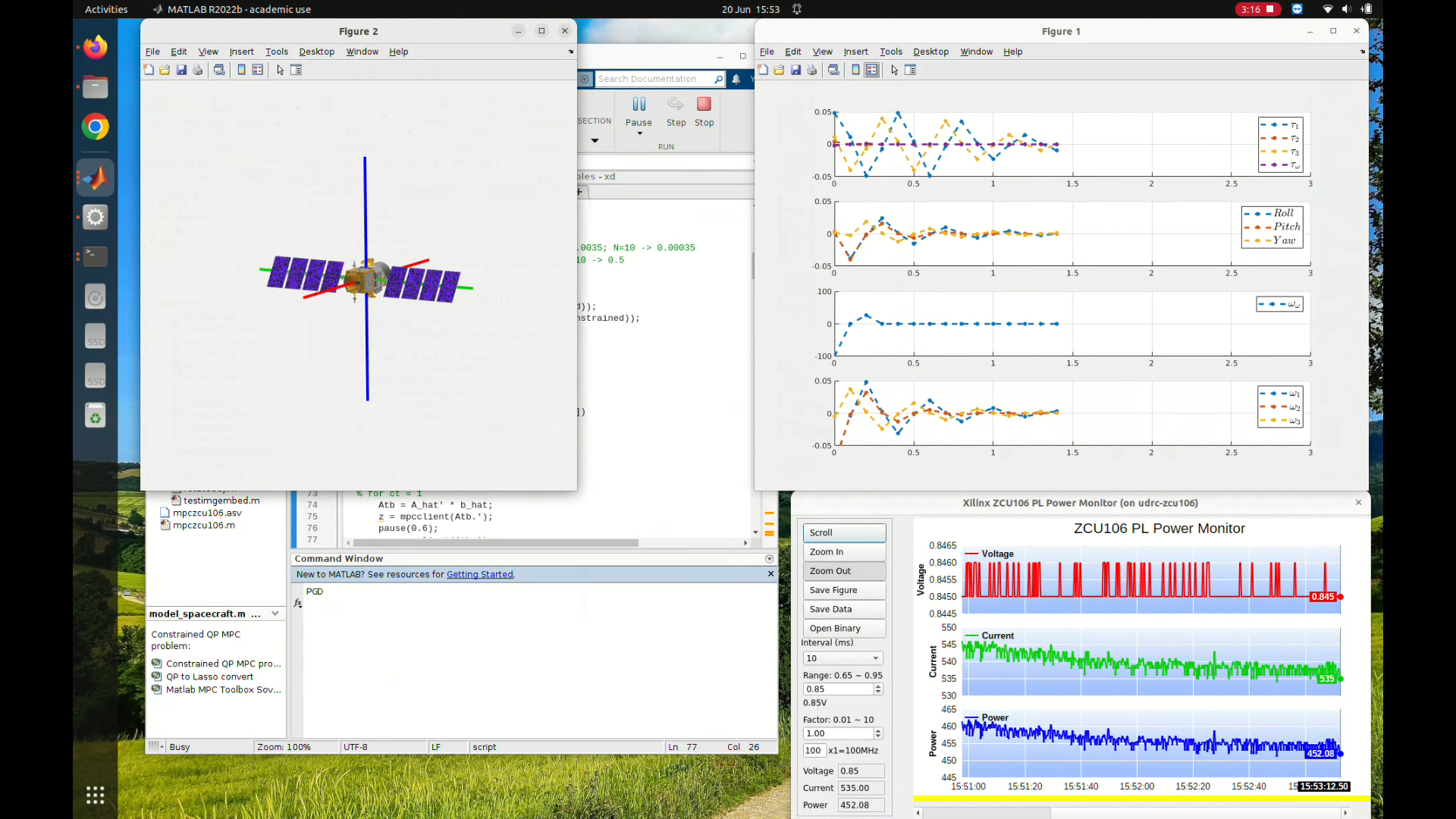}
        \caption{FPGA-In-the-Loop simulation of MPC with control horizon of $10$ samples, $64$ bit-width and $380$ mW of average power.}
        \label{Figure6.1}
        \end{figure}

Seven states are considered here:  Roll, Pitch, Yaw, $\omega_1$, $\omega_2$, $\omega_3$, $\omega_w$, where Roll, Pitch, Yaw describe the rotating angles of the body frame relative to the orbit  frame, and $\omega_1$, $\omega_2$, $\omega_3$ are the corresponding angular velocities. $\omega_w$ is the angular velocity along the spin axis. The thrusters are controlled by three input voltages, $\tau_1$, $\tau_2$, $\tau_3$, and the reaction wheel is controlled by input voltage $\tau_w$ accordingly.

\begin{figure}[!htb]
    \includegraphics[width=\columnwidth]{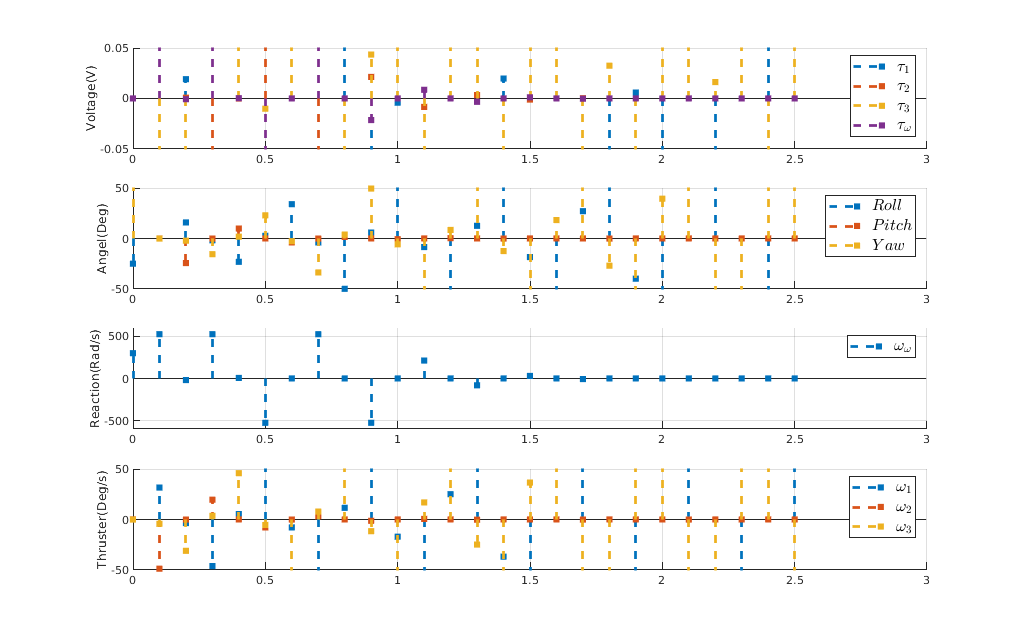}
    \caption{FPGA-In-the-Loop simulation of MPC with control horizon of $10$ samples, $28$ bit-width and $293$ mW of average power}
    \label{Figure6.2}
\end{figure}

In \autoref{Figure6.2}, we present an FPGA-In-the-Loop simulation of the MPC control for a fixed-point representation with a word length of 28 bits. It is observed that, while the average power consumption is marginally reduced to $293$ mW, this comes at the cost of growing oscillations around the equilibrium.

\begin{figure}[!htb]
    \includegraphics[width=\columnwidth]{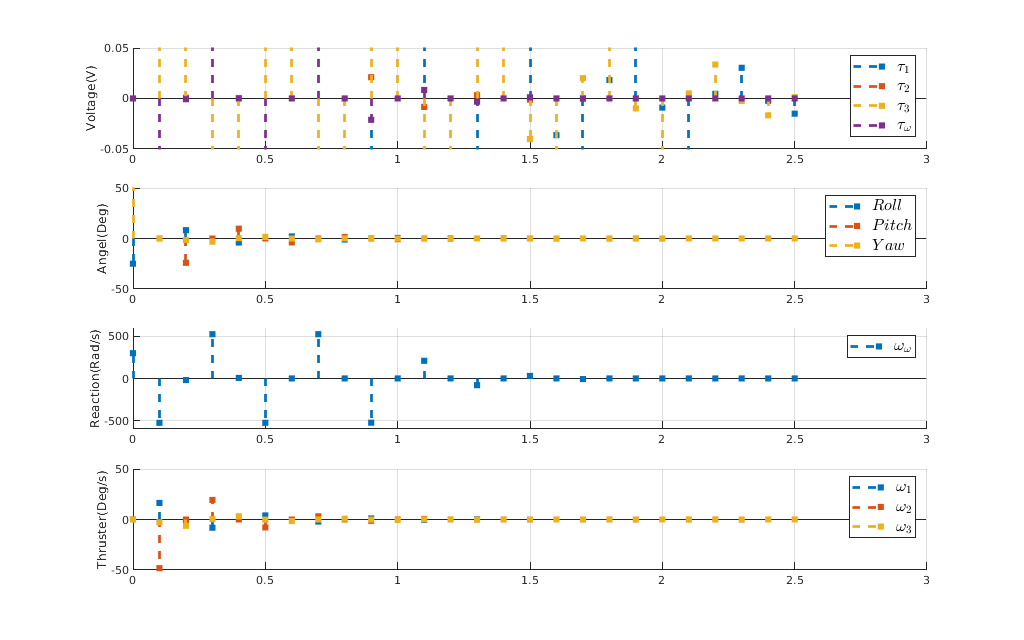}
    \caption{FPGA-In-the-Loop simulation of MPC with control horizon of $10$ samples, $W = 34$ bit-width, PL voltage of $0.85$ V and $298$ mW of average power.}
    \label{Figure6.3}
\end{figure}

\autoref{Figure6.3} offers a comparison with a 34-bit fixed-point representation. Operating at $0.85$ V PL voltage, this configuration consumes slightly more power ($298$ mW). Notably, this setting provides a better trade-off in terms of power and system stability as compared to the 28-bit version.

\begin{figure}[!htb]
    \includegraphics[width=\columnwidth]{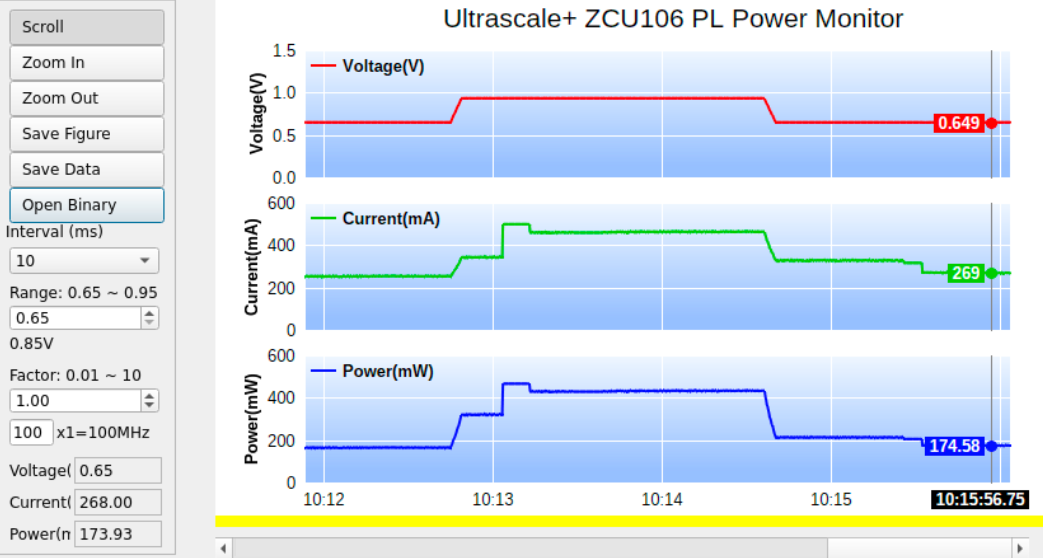}
    \caption{Real-time measurements of power, voltage, and current for 32 bits floating-point configuration.}
    \label{fig:third}
\end{figure}

\begin{figure}[!htb]
    \includegraphics[width=\columnwidth]{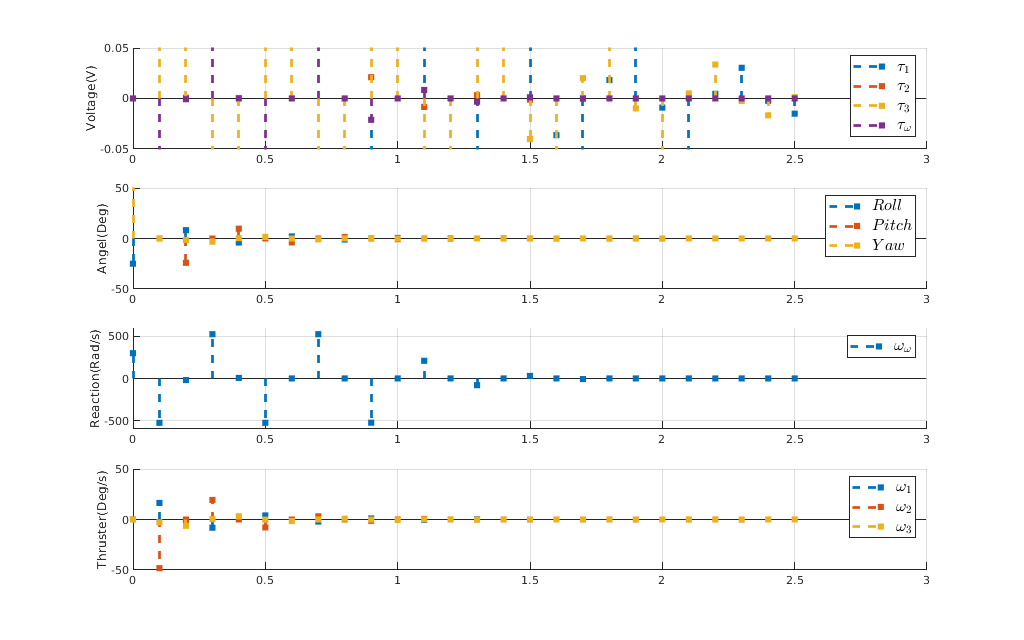}
    \caption{Simulated dynamics and control signals of FPGA-In-the-Loop MPC with 34 bits fixed-point configuration.}
    \label{fig:fourth}
\end{figure}

Switching to a 32-bit floating-point representation, \autoref{fig:third} visualizes the associated real-time measurements. The impact of this representation on the system dynamics can be observed in \autoref{fig:fourth}.

\begin{figure}[!htb]
    \includegraphics[width=\columnwidth]{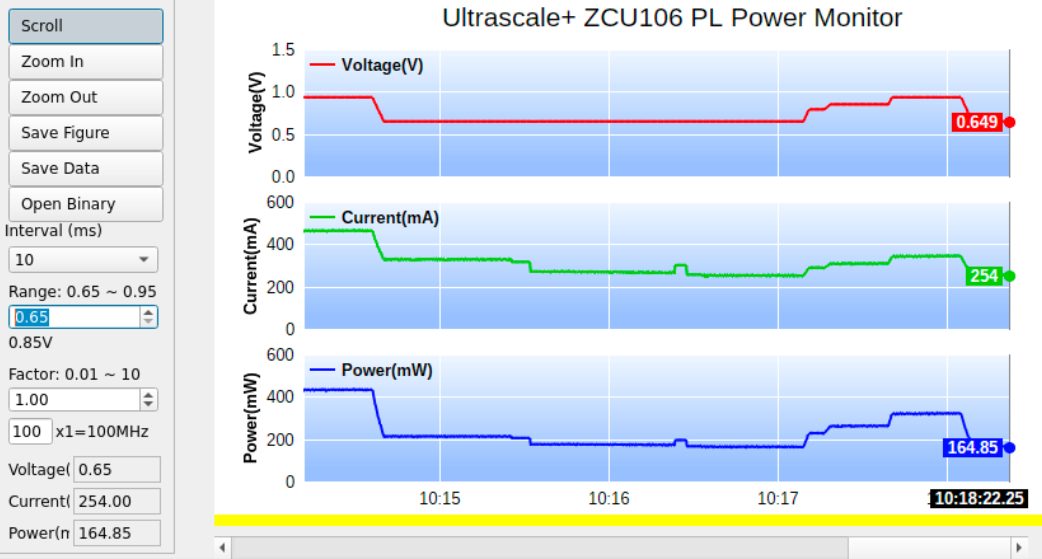}
    \caption{Real-time measurements of power, voltage, and current for PL voltages $V = 0.95$ and $V = 0.65$.}
    \label{fig:fifth}
\end{figure}

\begin{figure}[!htb]
    \includegraphics[width=\columnwidth]{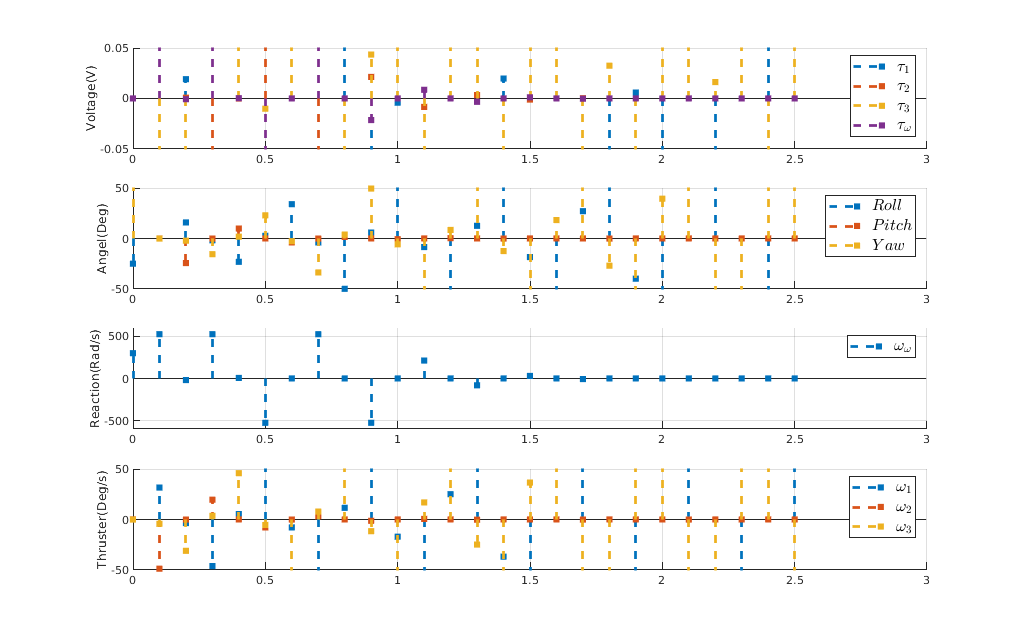}
    \caption{Simulated dynamics and control signals of FPGA-In-the-Loop MPC with 28 bits fixed-point configuration.}
    \label{fig:sixth}
\end{figure}

To explore the influence of varying the PL voltage, \autoref{fig:fifth} provides real-time measurements of power, voltage, and current for two distinct voltages ($0.95$ V and $0.65$ V). At the lower voltage of $0.65$ V, a 17\% power gain is realized when transitioning from double to 34 bits fixed-point representation. However, as highlighted in \autoref{fig:sixth}, further reduction in the number of bits introduces system instability, emphasizing the trade-offs to consider in selecting the right configuration for FPGA-based implementations.

\section{CONCLUSION}
\label{section:conclusion}
In this work, we have shed light on the critical role that meticulous bit-width selection plays in fixed-point representations, particularly when the aim is power-efficient FPGA implementations of MPC that do not sacrifice system stability. Through the versatile nature of FPGAs and by leveraging the inherent sparsity of the optimal control signal, we have demonstrated the superiority of reconfigurable hardware-centric first-order proximal gradient solvers. The proposed hardware-based control algorithm not only guarantee energy-conserving control solutions but also grant an unparalleled degree of control over the device's power usage, in this instance, an FPGA. Nevertheless, realizing the full potential of this methodology demands rigorous problem modeling, masterful hardware design, and comprehensive validation. As a future work, we will explore efficient implementations of nonlinear MPC on FPGA to further refine power consumption of the computing platform. 



\printbibliography
\end{document}